# Antichiral surface states and higher-order topological states based on a modified Haldane model

Jia-Rui Xu[1,2], Zhan Xiong[1,2,*], Kai Deng[1,2], Hai-Xiao Wang[3,†], Shiyang Liu[1,2], Yixian Qian[1,2], and Jian-Hua Jiang[4]

[1]*College of Physics and Electronic Information Engineering, Zhejiang Normal University, Jinhua 321004, China*

[2]*Key Laboratory of Optical Information Detecting and Display Technology of Zhejiang Province, Zhejiang Normal University, Jinhua 321004, China*

[3] *School of Physical Science and Technology, Ningbo University, Ningbo 315211, China*

[4]*Suzhou Institute for Advanced Research, University of Science and Technology of China, Suzhou 215123, China*

Correspondence and requests for materials should be addressed to xiongzhan@zjnu.edu.cn or wanghaixiao@nbu.edu.cn

Abstract: Antichiral surface states, characterized by unidirectional propagation on parallel surfaces, offer unique potential for controlling classical waves. However, their realization typically relies on complex implementations of the two-dimensional modified Haldane model, limiting practical applications. Here, we propose a simplified scheme to realize such states within the nodal-line semimetal phase of a single-layer honeycomb lattice, by emulating the essential physics of the modified Haldane model through an introduced layer degree of freedom. Furthermore, we demonstrate that unequal vertical interlayer couplings can generate valley higher-order topological partial bandgaps, hosting coexisting one-dimensional hinge states and gapped antichiral surface states. We numerically verify these multiple topological states in acoustic crystals, establishing a versatile platform for advanced wave manipulation.

## I. INTRODUCTION

Topological phases of matter [1,2] have attracted considerable attention owing to their rich topological phenomena and potential device applications. This interest was ignited by the discovery of the quantum Hall effect [3,4] under strong magnetic fields, which features chiral edge states that propagate unidirectionally along opposite sample edges. Building on this, the Haldane model [5] emerged as a pivotal theoretical model. This model, defined on a honeycomb lattice, introduces periodic local magnetic fluxes that cancel out on average, eliminating the need for a net external magnetic field while still generating chiral edge states. It not only establishes a framework for the Chern insulator but also lays the groundwork for exploring exotic topological phases in time-reversal-symmetry-invariant systems. Reversing the local fluxes on the sublattices modifies the Haldane model in a crucial way, yielding a distinct type of boundary propagation: the antichiral edge states [6]. Unlike chiral edge states, antichiral modes propagate in the same direction along both parallel edges, compensated by a counterflow in the bulk. This distinct bulk-boundary correspondence has sparked widespread interest [7-12] and inspired related concepts such as antihelical edge states [13]. However, the experimental realization of such antichiral edge states in electronic systems remains challenging, as it requires precisely engineered artificial gauge fields.

Classical wave metamaterials, such as photonic and phononic crystals, provide a versatile platform for realizing topological phenomena, benefiting from their macroscopic fabrication and tunability. To date, antichiral edge states have been realized using external magnetic fields in photonic systems [14-19], staggered air flows in acoustic systems [20,21], and intercomponent couplings in circuit lattices [22]. The unidirectional co-propagation of these states facilitates wave-manipulation effects such as wave splitting [21], paving the way for advanced acoustic wave control. More recently, three-dimensional (3D) layer-stacked systems have emerged as a powerful platform for engineering synthetic gauge fluxes. Implementations, such as time-reversal-breaking magnetic Weyl photonic crystals [23], time-reversal-invariant photonic semimetals [24], and acoustic square lattice systems [25], have successfully demonstrated two-dimensional (2D) antichiral surface states that feature co-propagating modes on opposing surfaces. However, a modified Haldane model that requires no external conditions has not been realized in acoustic systems.

Parallel to these developments, higher-order topology has risen to prominence, attracting widespread interest [26-36]. In such phases, the co-dimension of the topologically protected boundary states is equal to the difference between the system's bulk dimensionality and its order. Their physical origin can be understood through frameworks like Wannier center configurations [28,29,31,33] and the mass-domain picture of boundary geometries [27,30,36]. The higher-order topology is extended to topological semimetals, leading to higher-order topological semimetals [37]. These phases combine gapless bulk states characteristic of semimetals with lower-dimensional boundary modes protected by higher-order topology. They have been realized in classical wave systems and can be categorized as higher-order Weyl semimetals [37-50], higher-order Dirac semimetals [51-56], and higher-order nodal-line semimetals [57-63]. Notably, higher-order Weyl semimetals typically host one-dimensional (1D) hinge states alongside 2D chiral surface states. Although antichiral hinge states have been theoretically predicted [8] in composite Haldane systems, the relationship between the aforementioned 2D antichiral surface states and higher-order topology remains unexplored. Elucidating this connection could provide valuable insights for designing advanced multifunctional and multidimensional acoustic devices.

To fill these gaps, we propose two tight-binding models based on a modified Haldane model and leverage the layer degree of freedom to construct corresponding acoustic crystals. This approach bypasses the need for external fields or intricate 3D stacking. Our platform hosts symmetry-protected nodal lines coexisting with 2D antichiral surface states. Furthermore, building on this monolayer platform, the introduction of unequal vertical interlayer couplings splits these nodal lines into Weyl points and induces a higher-order topological phase with 1D hinge states protected by bulk polarization. Thus, our platform thereby achieves multidimensional topological control through the concurrent utilization of 2D antichiral surface transport and 1D hinge transport, providing a concrete pathway from surface to hinge physics in acoustic systems.

## II. MODEL I: ANTICHIRAL SURFACE STATES BASED ON A MODIFIED HALDANE MODEL

### A. Topological nodal-line semimetals and antichiral surface states in tight-binding models

The 3D nodal-line semimetal is constructed by stacking 2D single-layer honeycomb lattices directly on top of each other (A-A stacking) along the $z$ direction, as shown in Fig 1(a). The nearest-neighbor in-plane hopping and the next-nearest-neighbor interlayer hopping strengths are denoted as $t_0$ and $t_1$, respectively. A top view of the lattice and its hopping parameters is presented in Fig. 1(b). The orange and purple arrows denote the winding directions for the next-nearest-neighbor hoppings, which connect two A atoms or B atoms between adjacent layers (from lower to upper layers). The opposite circulation directions of these arrows indicate that the phases of the corresponding hopping terms have the same signs, which implies the two sublattice have opposite chirality. By utilizing the layer degree of freedom, this model realizes the next-nearest-neighbor hopping term $t_1 e^{-i\phi}$ of a modified Haldane model [6] through the interlayer coupling term $t_1 e^{-ik_z a_z}$. In the sublattice basis $(A, B)^T$, the Bloch Hamiltonian of the 3D tight-binding model can be expressed as:

$$H(\mathbf{k}) = \begin{pmatrix} h_{11} & h_{12} \\ h_{12}^* & h_{22} \end{pmatrix} \tag{1}$$

with $h_{11} = h_{22} = 2t_1 \left( 2\cos(-k_x a/2 + k_z a_z)\cos(k_y \sqrt{3} a/2) + \cos(k_x a + k_z a_z) \right)$, $h_{12} = t_0 \exp(-ik_y a/2\sqrt{3}) \left( \exp(ik_y a/\sqrt{3}) + 2\cos(k_x a/2) \right)$, where $\mathbf{k} = (k_x, k_y, k_z)$ is the Bloch wavevector. In the 2D honeycomb lattice, Dirac points are pinned at the corners of the 2D Brillouin zone. Consequently, the 3D unit cell is expected to host degenerate points at the hinges of the 3D Brillouin zone. The topological features of this model can be understood by examining the bulk dispersion (Fig. 1(d)) along the high-symmetry lines of the first Brillouin zone (Fig. 1(c)), where a twofold degenerate nodal line emerges along the $KH$ line. Figures 1(e) and 1(f) respectively show the projections of the 3D bulk dispersion around $(4\pi/3a, 0, 0.5\pi/a_z)$—the center of the $KH$ high-symmetry line—onto the $k_x k_y$ and $k_y k_z$ planes. In the $k_x k_y$ projection, the bulk bands exhibit a linear dispersion. These nodal lines, located at the Brillouin zone hinges, are depicted by the red lines in Fig. 1(c).

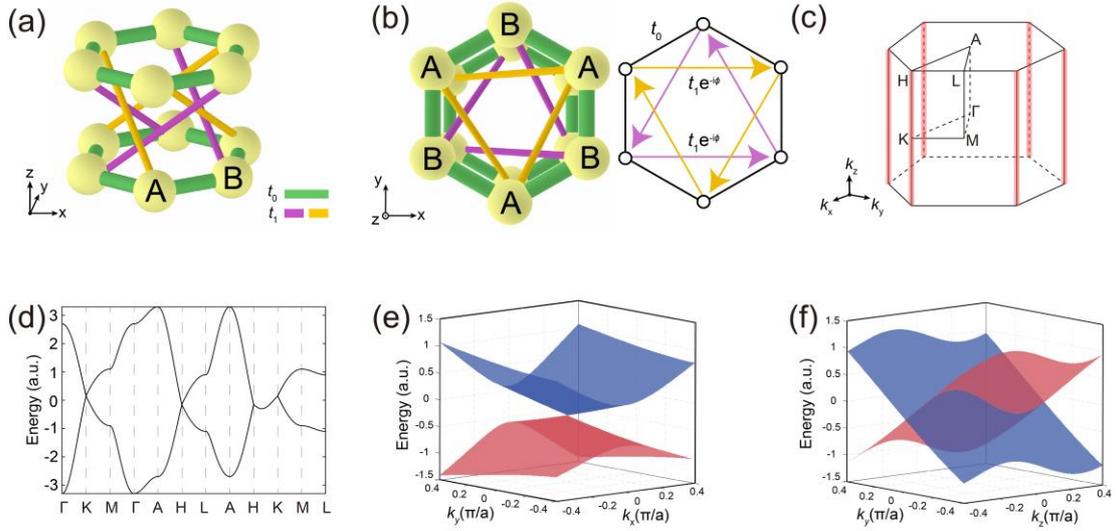

FIG. 1. (a) Schematic of single-layer unit cell of the honeycomb lattice with the antichiral interlayer couplings. (b) Top view of the antichiral interlayer couplings. The purple and orange arrows denote the opposite winding directions. The model is an analogy to the modified Haldane model. (c) The Brillouin zone of the honeycomb lattice shown in (a). (d) The bulk band structure of the unit cell along the high-symmetry lines. Band structures of the nodal-line semimetal phases near $(k_x, k_y, k_z) = (4\pi/3a, 0, 0.5\pi/a_z)$ projected onto (e) $k_x k_y$ and (f) $k_y k_z$ planes. The lattice parameters in the $xy$ plane and $z$ direction are $a = a_z = 1$. The coupling coefficients in the $xy$ plane and $z$ direction used in (d), (e), and (f) are $t_0 = -1$ and $t_1 = -0.05$, respectively.

To further confirm the nature of the $k_z$-dependent subsystems, we present the projected band structures for a ribbon strip with zigzag surfaces. The evolution of these band structures with varying $k_z$ is shown in Fig. 2. Evidently, the strip supports surface states (red and blue lines) across the range from $k_z = 0$ to $k_z = \pi/a_z$ even in the absence of a complete bulk band gap. Specifically, at $k_z = 0$, the interlayer coupling term $t_1 e^{-ik_z a_z}$ is a real number, the preservation of time-reversal symmetry ensures that the energies of the Dirac points at the $K$ and $K'$ points are equal, resulting in nearly flat surface-state bands. As $k_z$ increases, time-reversal symmetry breaking becomes stronger, causing the two Dirac points to shift in energy. Consequently, the surface states connecting the projections of the $K$ and $K'$ points are no longer flat and acquire a nonzero group velocity. Interestingly, the surface states from the upper and lower surfaces of the strip share an identical dispersion in momentum space. This indicates that the surface states on opposite surfaces possess the same chirality, which is a hallmark of antichiral surface states and is consistent with the results in Refs. [23,24]. At $k_z = \pi/a_z$, the band structures resemble those at $k_z = 0$.

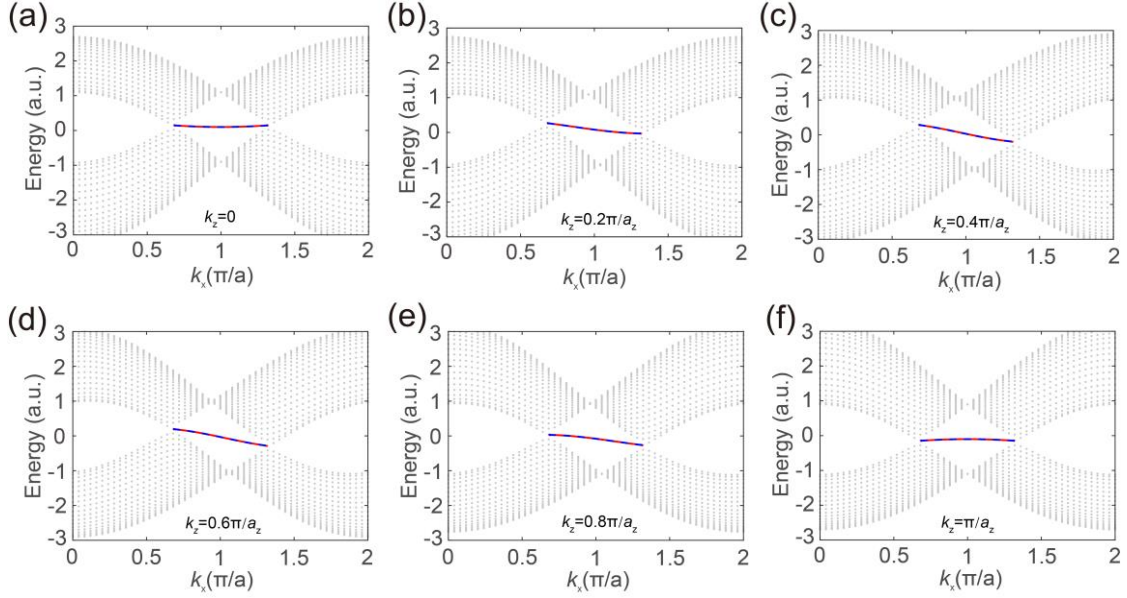

FIG. 2. Numerically calculated projected band structures of zigzag surface boundaries between unit cells and open boundary conditions for fixed values of (a) $k_z = 0$, (b) $k_z = 0.2\pi/a_z$, (c) $k_z = 0.4\pi/a_z$, (d) $k_z = 0.6\pi/a_z$, (e) $k_z = 0.8\pi/a_z$, and (f) $k_z = \pi/a_z$, respectively. The red and blue lines represent the antichiral surface states localized at the upper and lower surfaces of the ribbon supercell, respectively. The parameters are the same as in Fig. 1.

### B. Topological nodal-line semimetals and antichiral surface states in acoustic crystals

The above 3D tight-binding modified Haldane model can be realized in acoustic systems with coupled acoustic resonators and waveguides. If designed strictly according to the tight-binding model in Fig. 1(a), the three interlayer coupling waveguides with clockwise circulation and the three with counterclockwise winding would intersect in the corresponding acoustic model. This intersection contradicts the one-to-one correspondence of interlayer coupling coefficients in the tight-binding model. To ensure a faithful correspondence between the acoustic model and the tight-binding model, the three clockwise interlayer coupling waveguides (orange tubes) are shifted by $+1.4$ cm in the $z$-direction, while the three counterclockwise ones (purple tubes) are shifted by $-1.4$ cm. As shown in Fig. 3(a), the acoustic resonators and waveguides can be viewed as lattice sites and couplings in the tight-binding model. The interior of the unit cell is air-filled, with periodic boundary conditions along the periodic directions and acoustically rigid conditions on all other surfaces. Fig. 3(b) displays the acoustic bulk band structure, revealing a nodal line along the $KH$ path. This feature agrees with the tight-binding calculation in Fig. 1(d), confirming that the acoustic system successfully captures the distinct coupling coefficients of the tight-binding model.

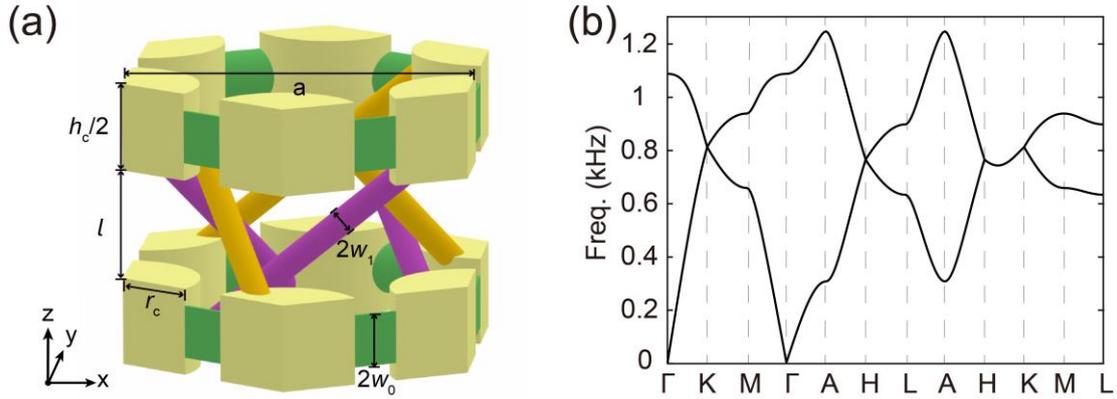

FIG. 3. (a) Side view of the single-layer unit cell of sonic crystals. The parameters of the unit cell are $a = 9\sqrt{3}$ cm, $r_c = 3.5$ cm, $h_c = 8$ cm, $w_0 = 1.2$ cm, $l = 5$ cm and $w_1 = 0.6$ cm. (b) Band dispersion of the unit cell along the high-symmetry lines of the Brillouin zone. The bands exhibit a nodal line along the $KH$ line.

The projected band structures for specific $k_z$ subsystems are displayed in Fig. 4, calculated using a slab-shaped acoustic supercell. The antichiral surface states (red and blue lines) are clearly doubly degenerate in every $k_z$ plane. Time-reversal symmetry dictates that the Dirac points at both $k_z = 0$ plane (Fig. 4(a)) and $k_z = \pi/a_z$ plane (Fig. 4(f)) share the same frequency. The band structures—including both bulk bands (gray lines) and antichiral surface bands—are symmetric with respect to $k_x = \pi/a$. At intermediate $k_z$ planes (Figs. 4(b)-4(e)), the Dirac points at the $K$ and $K'$ points undergo frequency splitting. This splitting causes antichiral surface bands connecting these points develop a non-zero group velocity. All these features are in excellent agreement with the results presented in Fig. 2.

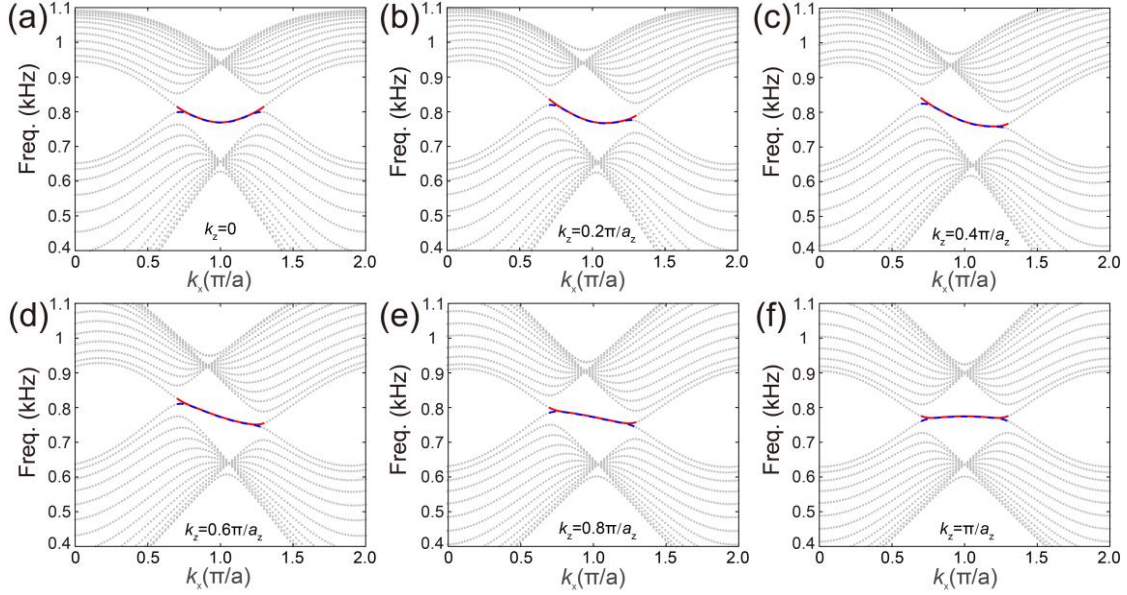

FIG. 4. (a-f) Evolution of the band structures of zigzag surface boundaries between acoustic ribbon supercell and hard boundaries for fixed values of $k_z = 0$, $k_z = 0.2\pi/a_z$, $k_z = 0.4\pi/a_z$, $k_z = 0.6\pi/a_z$, $k_z = 0.8\pi/a_z$, and $k_z = \pi/a_z$, respectively. The semi-infinite ribbon consists of 10 acoustic unit cells. Periodic boundaries are imposed in the $x$ and $z$ directions.

## III. MODEL II: HIGHER-ORDER TOPOLOGICAL STATES IN HIGHER-ORDER WEYL SEMIMETALS

### A. Higher-order Weyl semimetals and hinge states in tight-binding models

To realize the higher-order Weyl semimetal phase, we introduce unequal vertical interlayer couplings ($t_2 = -0.05$ and $t_3 = -1.25$ on the two sublattices) into the nodal-line semimetal model of Fig. 1. Side and top views of the resulting structure (termed the Type-B unit cell) are shown in Figs. 5(a) and 5(b), respectively. Correspondingly, we add the terms $2t_2\cos(k_z a_z)$ and $2t_3\cos(k_z a_z)$ to the $h_{11}$ and $h_{22}$ terms, respectively, in the Hamiltonian given by Equation (1). The bulk band structure is shown in Fig. 5(d). In contrast to the energy spectrum in Fig.1(d), the doubly degenerate nodal line along the $KH$ line collapses into a single Weyl point at the center of the $KH$ segment. Moreover, the nodal line within the range $k_z \in [-\pi/a_z, \pi/a_z]$ splits into two Weyl points in the $k_z = 0.5\pi/a_z$ and the $k_z = -0.5\pi/a_z$ planes, which carry topological charges of $+1$ and $-1$, respectively. A similar splitting occurs along the $K'H'$ path, yielding Weyl points with charges of $-1$ and $+1$ on the same $k_z$ planes. Consequently, the overall distribution of Weyl points in momentum space is summarized in Fig. 5(c). The linear dispersion around $(4\pi/3a, 0, 0.5\pi/a_z)$, as projected onto the $k_x k_y$ plane (Fig. 5(e)) and $k_y k_z$ plane (Fig. 5(f)), confirms the type-II nature of these Weyl points.

According to the distribution of Weyl points, the total topological charge in both the $k_z = 0.5\pi/a_z$ and $k_z = -0.5\pi/a_z$ slices is zero. Consequently, the Chern number associated with the partial bulk band gap vanishes for all constant-$k_z$ planes, except exactly at the planes hosting the Weyl points. This indicates that the insulating phases in these $k_z$ subsystems arises because sublattice-symmetry-breaking dominates over time-reversal-symmetry-breaking in opening the partial band gap [5,46]. Nevertheless, the partial band gaps in the two $k_z$ slices immediately above and below the Weyl points do undergo a gap closing and reopening process. Although these $k_z$ planes exhibit a zero Chern number, they are distinguished by their contrasting valley properties, namely, opposite and non-zero valley Chern numbers, signifying that a valley Hall phase transition occurs. Furthermore, they are characterized by different higher-order topological indices [33,46,64,65] of $\chi_{k_z=0} = (-1,0)$ and $\chi_{k_z=\pi/a_z} = (-1,1)$, respectively. More importantly, bulk polarizations and associated Wannier center positions can be utilized to further characterize the higher-order topology.

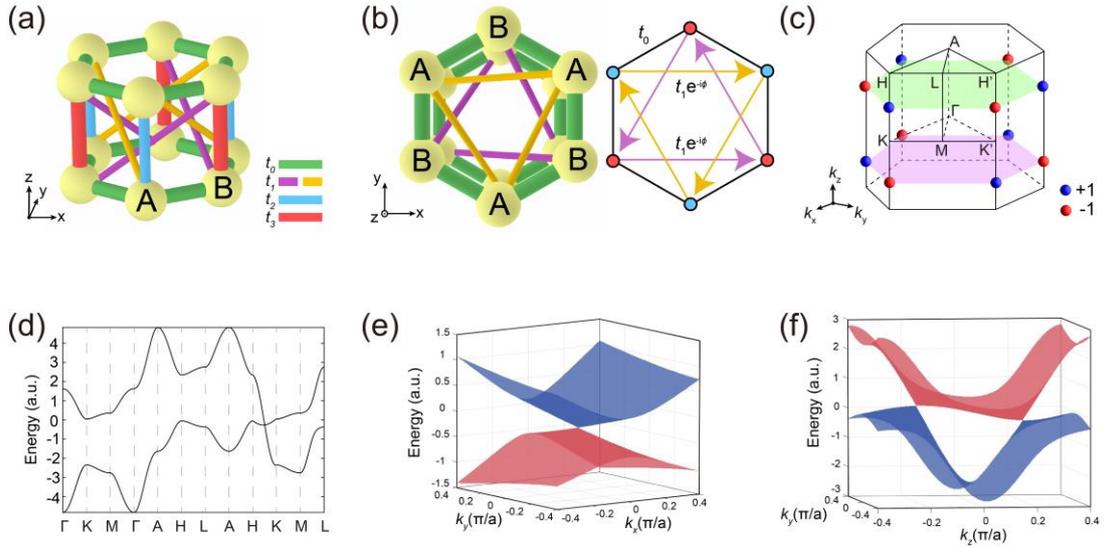

FIG. 5. (a) Schematic plot of single-layer unit cell of the honeycomb lattice with the two types of interlayer couplings: vertical and antichiral couplings. (b) Top view of the interlayer couplings. In addition to the antichiral interlayer couplings denoted by the purple and orange arrows, vertical couplings connecting the same lattice sites between adjacent layers are marked by red and blue points. (c) The Brillouin zone of the honeycomb lattice shown in (a). (d) Energy bands of the unit cell along the high-symmetry lines. Band structures of the higher-order Weyl semimetals near $(k_x, k_y, k_z) = (4\pi/3a, 0, 0.5\pi/a_z)$ projected onto (e) $k_x k_y$ and (f) $k_y k_z$ planes. The lattice parameters in the $xy$ plane and $z$ direction are $a = a_z = 1$. The coupling coefficients in the $xy$ plane and $z$ direction used in (d), (e), and (f) are $t_0 = -1$, $t_1 = -0.05$, $t_2 = -0.05$, and $t_3 = -1.25$, respectively.

According to the bulk-boundary correspondence, the nature of partial band gaps can be identified by the presence of surface states. Figure 6 shows the projected band structures of the corresponding 2D subsystems at fixed $k_z$ as $k_z$ is varied along the path of the bulk phase transition. One can clearly see that the antichiral surface states are gapped in the range $0 \leq k_z < 0.5\pi/a_z$, as exemplified in Figs. 6(a) and 6(b). This is in accord with the zero Chern number of these partial band gaps. At the critical plane $k_z = 0.5\pi/a_z$ (Fig. 6(c)), twofold degenerate antichiral surface states emerge, connecting the projections of Weyl points pinned at the $K$ and $K'$ points. For $k_z$ in the range $(0.5\pi/a_z, \pi/a_z]$, the antichiral surface states become gapped once more, as shown in Figs. 6(d)-6(f). This reopening of the surface gap results from sublattice-symmetry-breaking (governed by $|t_2 - t_3|$) playing a more significant role than time-reversal-symmetry-breaking (governed by $t_1$) in determining the partial band gaps.

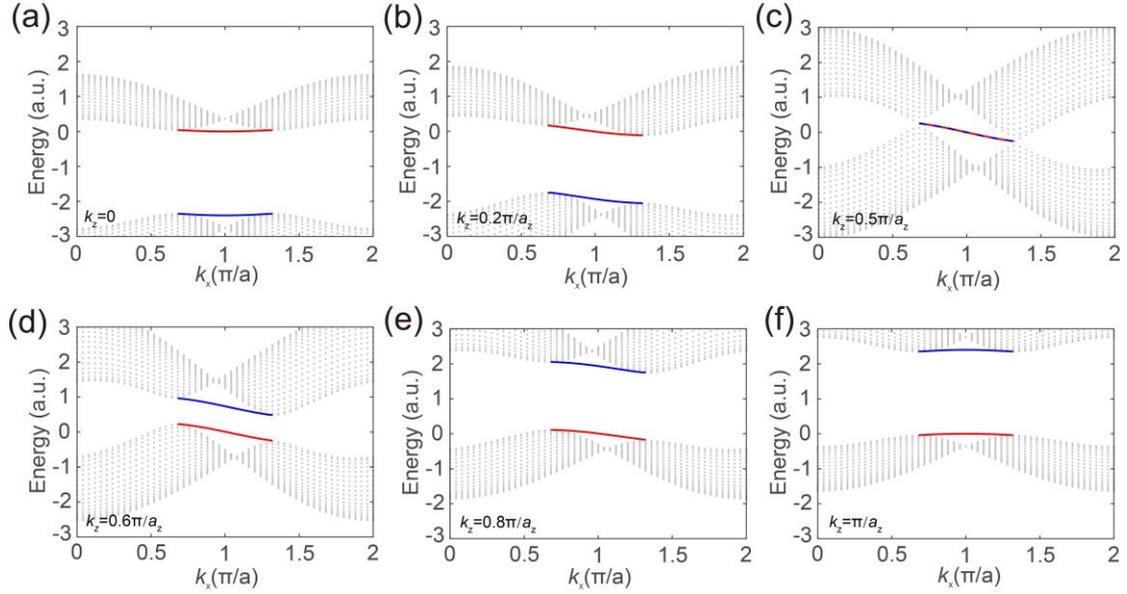

FIG. 6. Numerically calculated energy spectrum of zigzag surface boundaries between higher-order Weyl semimetal unit cells and open boundary conditions for fixed values of (a) $k_z = 0$, (b) $k_z = 0.2\pi/a_z$, (c) $k_z = 0.5\pi/a_z$, (d) $k_z = 0.6\pi/a_z$, (e) $k_z = 0.8\pi/a_z$, and (f) $k_z = \pi/a_z$, respectively.

We reveal the higher-order topology of the partial bulk band gaps by examining the 1D hinge states in a properly designed triangle-shaped supercell. As schematically shown in Fig. 7(a), the supercell has an outer side length of 12 Type-B higher-order Weyl semimetal unit cells and an inner side length of 6 Type-A higher-order Weyl semimetal unit cells, and is periodic along the $z$ direction. Specifically, the vertical coupling parameters ($t_2$, $t_3$) are swapped between the internal (Type-A) and external (Type-B) unit cells, while all other parameters remain identical. Consequently, the two regions, separated by the interface (i.e., three sides of the inner blue triangles) and hinges (i.e., three red solid circles), are both topologically trivial in terms of the Chern number ($C = 0$) but possess opposite valley Chern numbers across all $k_z$ subsystems except for $k_z = \pm 0.5\pi/a_z$. At these specific $k_z$ slices, the antichiral surface states become gapped, as shown in Fig. 7(b). Within these partial gaps, 1D hinge states emerge and are localized on the three distinct hinges, as indicated by the threefold degenerate red curve in the spectrum. Their emergence can be understood from the valley higher-order topological indices [46,64,65] and bulk polarization of the corresponding $k_z$ subsystems. The corresponding eigenfield distributions are displayed in Fig. 7(c) for the hinge states and in Fig. 7(d) for the gapped interface states. The realization of antichiral hinge states in our proposed system differs from that in the existing 3D model [8]. The latter consists of stacked 2D Haldane-model layers with opposite in-plane chirality, requiring a layer-dependent magnetization along the *z*-direction. In contrast, our system requires only single-layer unit cells that are periodic along *z*-direction and that together form a structure extended in the *xy*-plane.

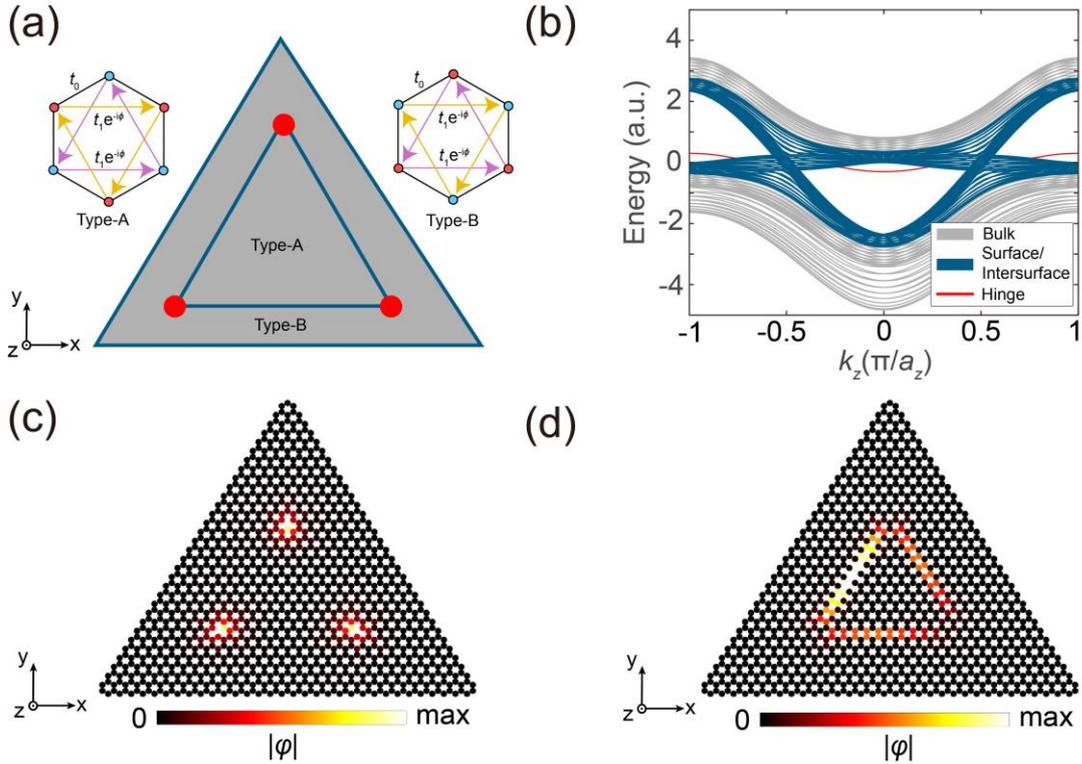

FIG. 7. (a) Schematic of an appropriately designed triangular prism supercell, configured to exhibit bulk, interface/surface, and hinge geometries. The interlayer couplings for the Type-A and Type-B higher-order Weyl semimetal unit cells are depicted in the insets. (b) Computed band diagram of the present model under the open boundary condition. In-gap hinge states denoted by red curves emerge for $0 < |k_z| < 0.239\,\pi/a_z$ and $0.761\pi/a_z < |k_z| < \pi/a_z$. Blue and gray zones show the interface/surface and bulk modes, respectively. Amplitude distributions of the (c) in-gap hinge and (d) interface

modes.

### B. Higher-order Weyl semimetals and hinge states in acoustic crystals

An acoustic higher-order Weyl semimetal is constructed based on the acoustic structure in Fig. 3(a). The $t_2$ and $t_3$ coupling coefficients of the tight-binding model (Fig. 5(a)) are realized by introducing vertical coupling waveguides with different radii of $w_2 = 0.6$ cm and $w_3 = 2$ cm on the two different sublattices. Figure 8(a) shows the side and top views of the unit cell, whose bulk band structure is presented in Fig. 8(b). A doubly degenerate Weyl point is found at $(4\pi/3a, 0, 0.292\pi/a_z)$ on the $KH$ path. Further analysis of the bulk bands in Fig. 8(d) reveals an additional Weyl point along the path ($k_x = 4\pi/3a$, $k_y = 0$), which is located at $k_z = -0.340\pi/a_z$. Similarly, along the path ($k_x = -4\pi/3a$, $k_y = 0$), two corresponding Weyl points are identified at $k_z = 0.340\pi/a_z$ and $k_z = -0.292\pi/a_z$. Consequently, there are a total of four Weyl points within the entire Brillouin zone, as depicted in Fig. 8(c).

Note that there is a minor difference in the Weyl point distributions between the acoustic unit cell and the tight-binding model: in the acoustic system, the Weyl points of opposite charges do not reside in the same $k_z$ slice. This leads to two separate regions (i.e., two orange regions in Fig. 8(d)) within the Brillouin zone possessing non-zero Chern numbers. More concretely, the Chern number is $+1$ for $k_z$ in the interval $(-0.34\pi/a_z, -0.292\pi/a_z)$ and $-1$ for $k_z$ in $(0.292\pi/a_z, 0.340\pi/a_z)$. In contrast, the Chern number is 0 in the following three intervals: $(-\pi/a_z, -0.34\pi/a_z)$, $(-0.29\pi/a_z, 0.29\pi/a_z)$, and $(0.34\pi/a_z, \pi/a_z)$.

As shown in Fig. 8(e), the eigenvalues of the $C_3$ rotation symmetry operator for the lowest acoustic bulk band are $e^{i0\pi}$ at the $\Gamma$ point and $e^{i\frac{2}{3}\pi}$ at the $K$ point. Correspondingly, at the A and $H$ points, the eigenvalues are $e^{i0\pi}$ and $e^{-i\frac{2}{3}\pi}$, respectively. These distinct eigenvalue distributions indicate that the topological phases of the two $k_z$ subsystems can be characterized by different valley higher-order topological indices [33]. Consequently, the higher-order topological indices are calculated as $\chi_{k_z=0} = (-1,0)$ for the $k_z = 0$ plane and $\chi_{k_z=\pi/a_z} = (-1,1)$ for the $k_z = \pi/a_z$ slice. This result agrees with the findings from the tight-binding model presented in Fig. 5. Moreover, the acoustic system exhibits a richer variety of topological phases along $k_z$, as illustrated by the phase transition from a valley higher-order topological phase to a Chern insulating phase shown in Fig. 8(d).

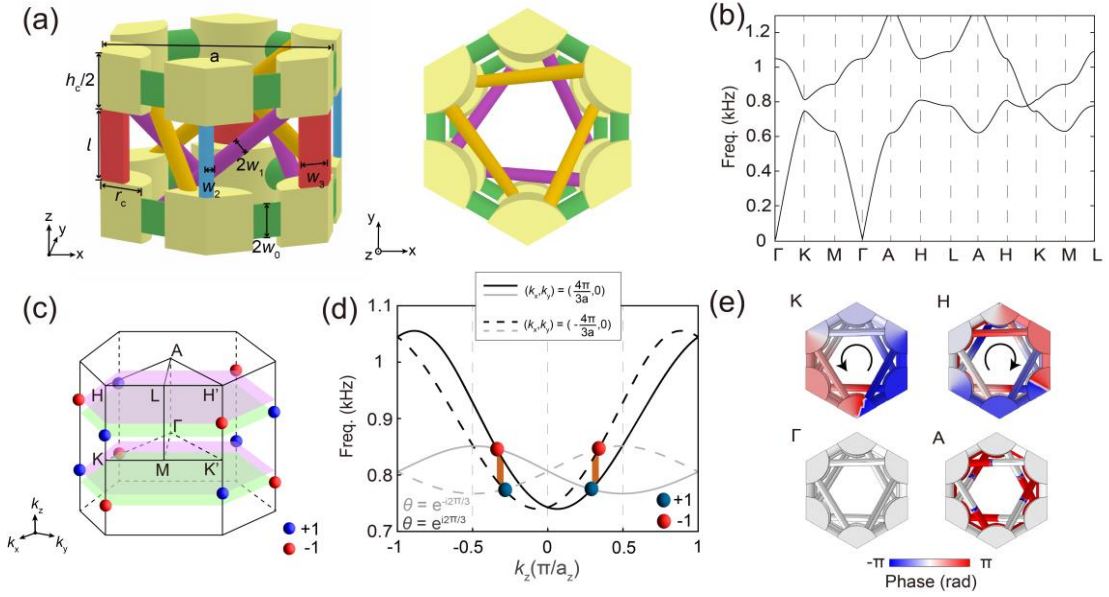

FIG. 8. (a) Side view of the single-layer unit cell of sonic crystals with two types of interlayer couplings. The geometric parameters of the unit cell are $r_c = 3.5$ cm, $h_c = 8$ cm, $l = 5$ cm, $w_0 = 1.2$ cm, $w_1 = 0.6$ cm, $w_2 = 0.6$ cm, and $w_3 = 2$ cm. (b) Band dispersion of the acoustic unit cell along the high-symmetry lines of the Brillouin zone. The bands have a Weyl point along $KH$ line. (c) The first Brillouin zone and Weyl points distribution. (d) Acoustic phase transitions from a Chern partial gap to a valley higher-order partial gap via a semimetal phase which hosts four type-II Weyl points. (e) Phase profiles for the acoustic eigenstates of the lowest bulk band at the $\Gamma$, K, A, and H points.

The acoustic projected band structures for different $k_z$ subsystems are shown in Fig. 9. At $k_z = 0$ (Fig. 9(a)), a band gap opens between the two antichiral surface bands which share identical group velocities. At $k_z = 0.292\pi/a_z$ (Fig. 9(b)), the Weyl point emerges at the $K$ point, where both these surface states and the bulk bands are degenerate. At the midplane $k_z = 0.315\pi/a_z$ between the two Weyl points (Fig. 9(c)), the surface states on the two opposite surfaces exhibit opposite chirality, a direct consequence of the nonzero Chern number of this slice. Similarly, at $k_z = 0.34\pi/a_z$ (Fig. 9(d)), Weyl point emerges at the $K'$ point, accompanied by the degeneracy of the antichiral surface states and bulk bands at this point. As $k_z$ increases further to $0.5\pi/a_z$ and $1.0\pi/a_z$ (Figs. 9(e) and 9(f)), a band gap reopens between the two antichiral surface bands. While the specific $k_z$ values at which the antichiral surface bands become degenerate differ between the tight-binding model and the acoustic system—due to the difference in Weyl point distribution—these degeneracies consistently correspond to the $k_z$ slices hosting the Weyl points. Despite this difference, the key features of the tight-binding model results (Fig. 6) are well reproduced in the numerically computed projected acoustic bands.

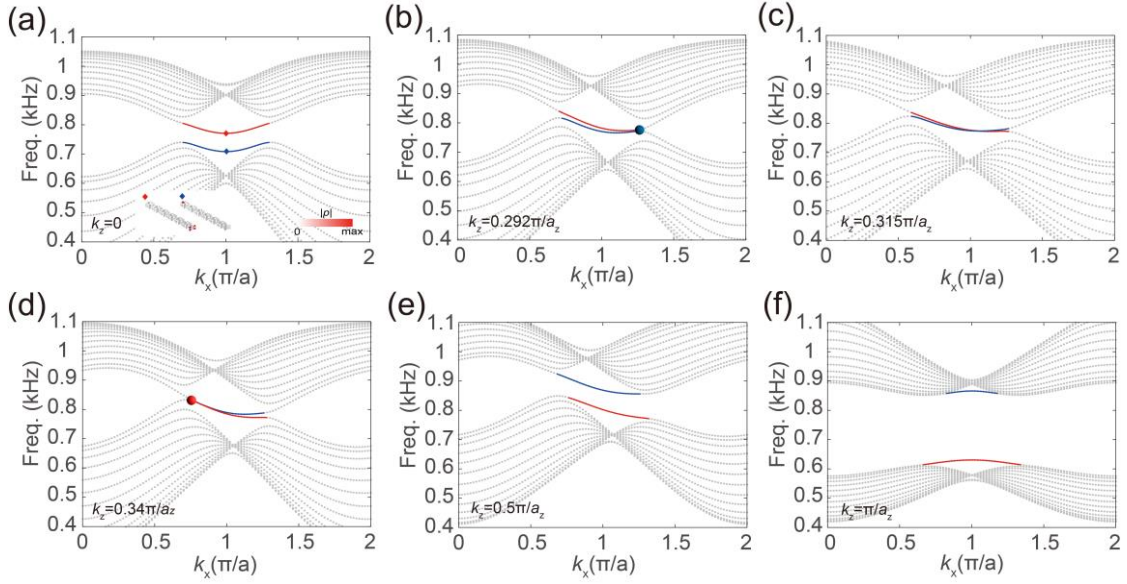

FIG. 9. (a-f) Projected band structures of zigzag surface boundaries between acoustic ribbon supercell and hard boundaries for 2D $k_z$ subsystems of $k_z = 0$, $k_z = 0.292\pi/a_z$, $k_z = 0.315\pi/a_z$, $k_z = 0.34\pi/a_z$, $k_z = 0.5\pi/a_z$, and $k_z = \pi/a_z$, respectively.

The supercell structure for the acoustic system is constructed according to the schematic in Fig. 7(a), with acoustically hard boundary conditions in the $xy$-plane and periodic boundary conditions along the $z$ direction. The acoustic projected bands are shown in Fig. 10(a), where the blue and red regions correspond to surface/interface states and hinge states, respectively. In contrast to the tight-binding results in Fig. 7(b), the hinge bands near $k_z = 0$ are obscured by the surface/interface states, whereas well-defined hinge states emerge only near $k_z = \pm\pi/a_z$. Around $k_z = 0$, the surface band gap is relatively small, while around $k_z = \pm\pi/a_z$ it is significantly larger. This larger surface band gap promotes the localization of hinge states by suppressing their hybridization with the surface states. The acoustic pressure field distributions at $k_z = 0.9\pi/a_z$ are presented for a representative hinge state ($f = 0.85056$ kHz) in Fig. 10(b), a surface state ($f = 0.81579$ kHz) in Fig. 10(c), and an interface state ($f = 0.96304$ kHz) in Fig. 10(d).

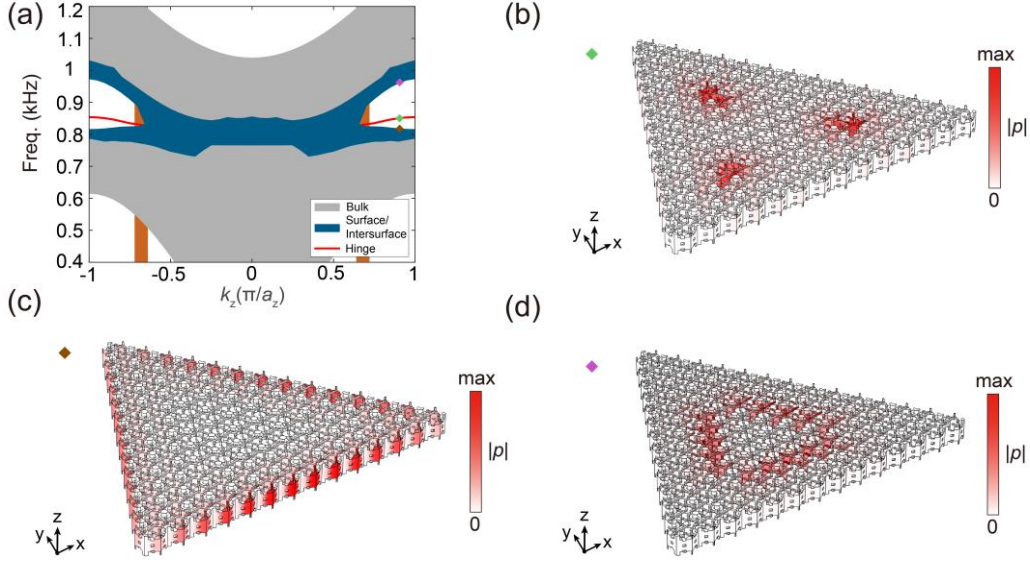

FIG. 10. (a) Numerically calculated band diagram of the properly designed acoustic triangular prism supercell. Three degenerate in-gap hinge states denoted by red curves emerge for $0.7\pi/a_z < |k_z| < \pi/a_z$. Simulated acoustic pressure distributions of the (b) hinge, (c) surface, and (d) interface states.

## IV. CONCLUSIONS AND DISCUSSIONS

In summary, we have realized a modified Haldane model in a single-layer honeycomb acoustic lattice through engineered interlayer couplings. The resulting system provides a simplified platform for exploring exotic topological states. Our key findings are threefold. First, with only antichiral interlayer coupling, the system exhibits symmetry-protected nodal lines and hosts 2D antichiral surface states within each constant-$k_z$ subsystem. Second, introducing unequal vertical couplings splits these nodal lines into Weyl points, which serve as phase boundaries separating partial bandgaps that exhibit valley-dependent higher-order topology. Third, within these bandgaps, we find robust 1D hinge states, whereas the antichiral surface states existing in other $k_z$ slices become gapped. All theoretical predictions, based on a sequence of two tight-binding models, are numerically verified in our designed acoustic crystals. This work not only establishes a direct link between antichiral surface transport and higher-order topology but also demonstrates concurrent multidimensional wave control by leveraging both 2D surface and 1D hinge states. Our platform thus opens avenues for designing advanced acoustic devices, such as wave splitters, topological sensors, and robust waveguides.

## ACKNOWLEDGMENTS

This work was supported by the National Key Research and Development Program of China (Grant No. 2022YFA1404400), the National Natural Science Foundation of China (Grants No. 12125504, No. 12074281, No. 12204417, No. 12474432, and No. 12474301), the "Hundred Talents Program" of the Chinese Academy of Sciences, the Zhejiang Provincial Natural Science Foundation of China (Grant No. LQ22A040005), and the Priority Academic Program Development of Jiangsu Higher Education Institutions.